\documentclass[pre,twocolumn,showpacs]{revtex4}
\usepackage{graphics,amsmath,amssymb}
\usepackage{dcolumn}

\begin{document}
\title{Login times in e-mail servers are scale-free}

\author{Arnab Chatterjee}
\email{arnab@cmp.saha.ernet.in}
\affiliation{Saha Institute of Nuclear Physics,  Block-AF, Sector-I Bidhannagar, Kolkata-700064, India.}

\begin{abstract}
The login time distribution in e-mail servers are studied.
The distribution seems to be decaying in a scale-free manner with a decay
exponent lying between 1.2 and 2.2, varying from server to server. A simple 
mean-field theory shows that this distribution has a decay exponent 2.
\end{abstract}

\pacs{89.20.Hh,89.75.Hc,89.75.Da,43.38.Si}
\maketitle

\noindent
Internet has drastically changed the mode of communication in the last few 
decades. E-mail has made communication efficient, fast and easy across the
globe and helps business, scientific and social interactions.
The structure of this communication network is dynamic, where 
each user is a node and the packet of information trasmitted (an e-mail)
constitutes a link between a pair of nodes (users, as sender and recipient).
The structure of such a network has been found to be scale-free \cite{ebel}
and is somewhat similar to network structure of the internet \cite{alb1}. 
E-mail has been looked upon as a social network \cite{well}, whose degree 
distribution and mode of information flow has also been studied \cite{info}. 
Given the network topology, how the individual nodes respond temporally in 
terms of their activity has not yet been studied. A very important and 
immediate example in this respect is a study of lifetimes of different logins 
in an e-mail server. The fundamental question stands as: How much time does 
an user spend during a particular login in his/her e-mail account? It can 
range from very small time-scales (bounded by the time required to log out 
immediately after logging in) to practically infinity. But is practice, 
it ranges from the order of seconds to many days, depending on the purpose 
and necessity of the login. 

Analyzing data from stock-market \cite{bbm1}, Internet (number of links to 
a page \cite{alb1}, number of pages within a site \cite{ah1}),
as well as natural phenomena like earthquake \cite{gut}, rainfall \cite{rain} 
etc. yielded interesting observations, and our study on the distribution of 
login times is not an exception. Login time data from different countries 
are distributed in time according to an universal power law: there are many 
logins of short durations, whereas, very few extend for very large times. 
This universal distribution can be explained by a simple mean-field theory.

\begin{figure}
{\centering \resizebox*{8cm}{7.2cm}{\includegraphics{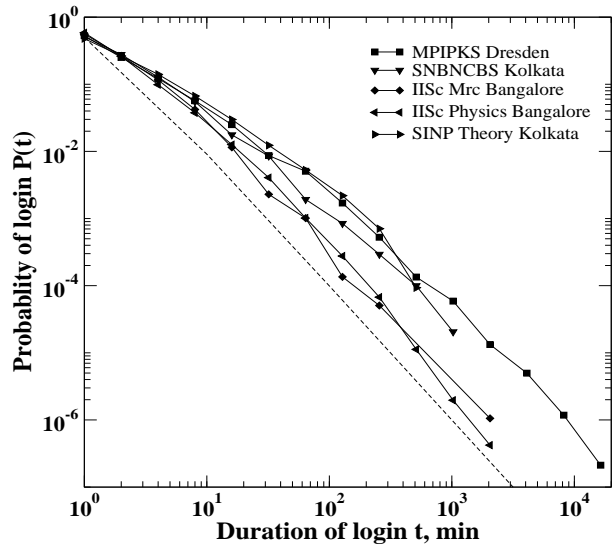}} \par}
\caption{\label{fig1} The probability distribution $P(t)$ of logins at 
different login times $t$ in minutes for some e-mail servers. The data has
been binned logarithmically to reduce noise. The decay exponent varies 
between $1.2$ to $2.2$ (see Table~\ref{tab:table1}). The dotted line is
the mean-field solution (Eqn. \ref{cumu1}).}
\end{figure}

We collected login time data from e-mail servers of all major 
research centres in India and a few places in different parts of the world. The 
computers register login times to the extent of detailed particulars as
user, date, login entry time, login exit time and net login time. The 
registered login times are in units of minutes, as it is the standard
time resolution set to measure login times in the machines. Hence, all data 
collected were in terms of integral minutes. The login time distribution
(see Fig~\ref{fig1}) turned out to be scale-free, i.e,
\begin{equation}
\label{emp}
P(t) \sim t^{-\alpha}
\end{equation}
with the decay exponent $\alpha$ varying between 1.2 to 2.2 from server to 
server. Table~\ref{tab:table1} shows the decay exponents for different 
e-mail servers.

The data suggest that there are more logins of shorter durations than of
longer durations, which is quite natural. Speculations could only
predict the decaying nature of the distribution, but not why it is scale-free.
The data shows that there is no typical time-scale of a login.

\begin{table}
\caption{\label{tab:table1} The details of the empirical observations: Name 
of E-mail servers and their approximate decay exponents.}

\begin{ruledtabular}
\begin{tabular}{lc}
{E-mail server} & {Decay exponent}\\
\hline
{MPIPKS Dresden} & {1.8}\\
{IISc Mrc, Bangalore} & {1.9}\\
{IISc Physics, Bangalore} & {2.2}\\
{SNBNCBS Kolkata} & {1.5}\\
{IMSc Chennai} & {1.3}\\
{SINP Theory} & {1.2}\\
{SINP Cmp} & {1.6}\\
{SINP Surf} & {1.7}\\
{SINP Lotus} & {1.5}\\
{SINP Petal} & {1.5}\\
\end{tabular}
\end{ruledtabular}
\end{table}

The origin of different login times are due to human behavior to check e-mails
frequently between work, while those of larger time scales are due to logins
which take place once or twice in a day to check e-mails coming in the previous
few hours. The even larger time scales arise due to file-transfer processes,
downloads and even due to reasons like `carelessness'!

We explain the scale-free nature of the login times by a simple mean field
theory. We define $p$ as the probability of any user to log out at any time. 
We, however, consider no interaction between users for simplicity, and 
effects of heterogeneity and frustration are only realized by assigning 
a distribution of $p$. Hence, we treat this as a single-particle problem, 
and the cumulative effect in the many-particle situation is realized by 
assuming a distribution of $p$.

As per our assumption, the probability of an user to have not logged out at 
time $t-1$ is $(1-p)^{t-1}$ and logs out at time $t$ is
$\rho(p,t)=p(1-p)^{t-1}$. For a multi-user system, where $p$ follows a 
distribution $f(p)$, the probability of logging out at time $t$ is 
\begin{equation}
\label{cumu}
P(t) = \int_0^1 \rho(p,t) f(p) dp.
\end{equation}

\noindent
We assume that the distribution of $p$ among the users is fairly uniform, i.e,
$f(p)=1$, which gives, 
\begin{equation}
\label{cumu1}
P(t) = \frac{1}{t(t+1)}
\end{equation}

\noindent
which in asymptotic limit gives $P(t) \sim t^{-2}$. Fig~\ref{fig1} shows a
comparison of the real data and Eqn.(\ref{cumu1}).

We have studied the empirical distribution of login times in e-mail servers.
This observed behavior is unique for e-mail servers and 
differ from general purpose computers and computational servers, which show 
a peaked distribution with an exponential (or stretched exponential) decay. 
The distribution of login times in e-mail servers, however, also has a peak 
which is not observed due to the fact that the minimum time resolution there 
is greater than the typical time-scale at which this peak occurs. The origin 
of the observed distribution is but a result of human behavior. Human behavior 
giving rise to frustration and fluctuations have been approximated by a 
simple mean-field theory. Our mean-field result is compatible with the real 
data and can reasonably explain the origin of scale-free nature of login time 
distribution. Apart from the e-mail systems studied, this interesting
behavior of a large number of components may also lead to similar observations
in systems with similar dynamical topology as in other modes of social
communication like telephone network and verbal conversation.

\begin{acknowledgments}
The author acknowledges B. K. Chakrabarti, P. Bhattacharyya for discussions 
and K. Das for technical assistance with the data. The author also 
acknowledges \cite{datasrc} H. Scherrer and A. K. Chattopadhyay for data
from MPIPKS Dresden, P. A. Sreeram for data from SNBNCBS, S. Chattopadhyay,
A. DeSarkar, B. Muthukumar and I. Sarkar and System Administrators of SINP
e-mail servers for data from SINP, 
P. Ray and G. Subramoniam for data from IMSc, Chennai, V. Shenoy and C. Karthik
for data from IISc-MRC \& Srinivas V. C. for data from IISc-Physics, Bangalore.
\end{acknowledgments}


\end{document}